\documentclass[aps,prb,twocolumn,superscriptaddress]{revtex4-2}

\usepackage{graphicx}
\usepackage{dcolumn}
\usepackage{bm}
\usepackage[colorlinks=true,allcolors=blue]{hyperref}
\usepackage{color}

\begin{document}

\title{Quasibound states in the continuum\\ in photonic-crystal-based optomechanical microcavities}

\author{Cindy Péralle}
\affiliation{Department of Physics, Chalmers University of Technology, SE-412 96 Göteborg, Sweden}

\author{Sushanth Kini Manjeshwar}
\affiliation{Department of Microtechnology and Nanoscience, Chalmers University of Technology, SE-412 96 Göteborg, Sweden}

\author{Anastasiia Ciers}
\affiliation{Department of Microtechnology and Nanoscience, Chalmers University of Technology, SE-412 96 Göteborg, Sweden}

\author{Witlef Wieczorek}
\affiliation{Department of Microtechnology and Nanoscience, Chalmers University of Technology, SE-412 96 Göteborg, Sweden}

\author{Philippe Tassin}
\affiliation{Department of Physics, Chalmers University of Technology, SE-412 96 Göteborg, Sweden}

\date{\today}

\begin{abstract}
We present a detailed study of mechanically compliant, photonic-crystal-based microcavities featuring a quasibound state in the continuum. Such systems were recently predicted to reduce the optical loss in Fabry-Pérot-type optomechanical cavities. However, they require two identical photonic-crystal slabs facing each other, which poses a considerable challenge for experimental implementation. We investigate how such an ideal system can be simplified and still exhibit a quasibound state in the continuum. We find that a suspended photonic-crystal slab facing a distributed Bragg reflector realizes an optomechanical system with a quasibound state in the continuum. In this system, the radiative cavity loss can be eliminated to the extent that the cavity loss is dominated by dissipative loss originating from material absorption only. These proposed optomechanical cavity designs are predicted to feature optical quality factors in excess of $10^5$.
\end{abstract}

\maketitle

\section{Introduction}

Reducing optical loss is paramount for a variety of engineered devices. Loss of confined modes can be reduced by designing structured materials that create bandgaps around the modes of interest~\cite{akahane2003high,englund2005general,deotare2009high}. An alternative strategy is to use a bound state in the continuum (BIC)---a nonradiating localized mode decoupled from the continuum of modes~\cite{tassin2012,hsu2016,koshelev2019nonradiating,azzam2021photonic,huang2023}. In theory, their infinite quality factor make BICs particularly interesting for efficient light confinement with applications in filtering~\cite{foley2014symmetry}, lasing~\cite{kodigala2017lasing,wu2020room}, and sensing~\cite{liu2017optical,romano2018optical}. In practice, losses due to material absorption, finite sample size, and structural disorder limit the achievable quality factor. No longer fully decoupled from the external radiation, the BIC then becomes a quasi-BIC with a high, yet finite quality factor that can be experimentally observed.

Recently, the concept of BICs was applied to reducing loss in cavity optomechanical devices \cite{Aspelmeyer2014}, independently for the mechanical mode~\cite{liu2022optomechanical,zhao2019mechanical} and the optical mode \cite{Fitzgerald2021}. Losses in optomechanical systems limit the achievable displacement sensitivity, the coherent interaction strength between the mechanical resonator and the resonant electromagnetic field, and the lifetime of optomechanical quantum states. Mechanical loss can be drastically reduced using a wide variety of methods including phononic bandgap engineering, strain engineering, soft clamping techniques, or inverse design~\cite{yu2014phononic,tsaturyan2017ultracoherent,ghadimi2018elastic,hoj2021ultra}. However, optical loss is still a major roadblock for achieving, for example, the regime of single-photon strong optomechanical coupling~\cite{rabl2011photon,nunnenkamp2011single}.

\begin{figure}[b]
    \includegraphics{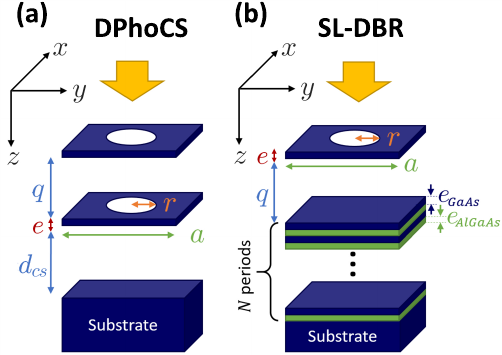}
    \caption{\label{syst} Schematic representation of a unit cell of the DPhoCS (a) and SL-DBR (b) systems. (a)~DPhoCS: a cavity (length $q$) formed by two PhC slabs made of cylindrical holes (radius $r$) arranged in a square lattice (lattice constant $a$) in a GaAs slab of thickness $e$. The cavity is suspended above a GaAs substrate at a distance $d_\mathrm{cs}$. (b)~SL-DBR: a cavity (length $q$) formed by a PhC slab made of cylindrical holes (radius $r$) arranged in a square lattice (lattice constant $a$) surrounded by GaAs, and a DBR with $N$ pairs of alternating GaAs and AlGaAs layers of thicknesses $e_{GaAs}$ and $e_{AlGaAs}$, respectively. This DBR is grown on a GaAs substrate.}
\end{figure}

In this article, we focus on analyzing practical optomechanical setups that realize optical quasi-BICs with photonic crystal (PhC)-based optomechanical microcavities. Suspended PhCs combine excellent mechanical and optical properties due to their small weight, low mechanical dissipation, and engineerable optical reflectivity~\cite{bui2012high,makles20152d,norte2016mechanical,kinimanjeshwarSuspendedPhotonicCrystal2020,zhouCavityOptomechanicalBistability2022,manjeshwarMicromechanicalHighQTrampoline2022}. As a consequence, PhC slabs have been considered in theoretical proposals for optomechanical systems \cite{naesbyMicrocavitiesSuspendedSubwavelength2018,cernotikCavityQuantumElectrodynamics2019} and have been used in various optomechanical experiments already~\cite{gartner2018integrated,jongCoherentMechanicalNoise2022,zhouCavityOptomechanicalBistability2022,enzian2022phononically}. Recently, it was shown that a cavity optomechanical system formed between two suspended PhC slabs realizes a Fabry-Perot-type optical BIC \cite{hsu2016,huang2023}, exhibiting a quality factor only limited by intrinsic loss in the material \cite{suh2003displacement,suh2005displacement,marinica2008bound,Fitzgerald2021}. Here, we extend the analysis of Ref.~\cite{Fitzgerald2021} to setups that are more amenable to experimental realization and feature a quasi-BIC, which can be accessed by external radiation, as required for interfacing with optomechanical devices.

\section{Two photonic-crystal membranes}

\subsection{Model description} \label{sec: Model description}

Many device structures based on BICs presented in the literature exploit a few known geometries: photonic crystals that exhibit symmetry-protected BICs at the $\Gamma$ point \cite{cong2019symmetry,li2019symmetry} or Friedrich-Wintgen BICs in one-dimensional cavities \cite{amrani2022friedrich,lee2020bound,azzam2018formation}. These high-symmetry geometries are often incompatible with planar fabrication techniques. Here, we start from the double photonic-crystal-membrane cavity (DPhoC) as proposed in Ref.~\cite{Fitzgerald2021}, and adapt it to allow for its fabrication on a substrate. This cavity is formed by two identical suspended PhC slabs (thickness $e$), separated by a distance $q$ and patterned with cylindrical holes in a square lattice (radius $r$, lattice constant $a$). We first study a PhC with this square lattice, but also look into a hexagonal pattern at a later stage (Sec.~\ref{Section_hexa}).

In practice, the two photonic-crystal membranes are suspended over a substrate with a distance $d_{cs}$ between the substrate and the first PhC membrane~\cite{kinimanjeshwarSuspendedPhotonicCrystal2020}. The resulting system (DPhoCS), depicted in Fig.~\ref{syst}(a), is excited with a normally incident light beam with wavelength $\lambda_0$ (frequency $f_0$). Focusing on a frequency interval typical for tunable lasers in the telecom range, from about 184 to 197\,THz, we search for bound states in the continuum in this frequency interval and for the case in which the suspended PhC membranes and the substrate are made from GaAs. We note that the same approach can be used for finding bound states in the continuum at other frequencies and with other materials, for example with SiN \cite{gartner2018integrated} or InGaP \cite{manjeshwarMicromechanicalHighQTrampoline2022}. In the spectral range around 190\,THz, GaAs can be modeled as a medium with complex refractive index $n= n_0 + i n_I$, where $n_0 = 3.374$ and $n_I=4.4 \times10^{-6}$~\cite{Fitzgerald2021,michael2007wavelength,xu2009influence}.

All results in this article are obtained by calculating the eigenmodes of the optical structures using a finite-element package (COMSOL), where we model a unit cell of the periodic structure with periodic boundary conditions. This model does not take into account the finite dimensions of the actual system, possibly leading to a lower PhC reflectivity \cite{toft-vandborgCollimationFinitesizeEffects2021}. Moreover, another simplification is made on the incident beam, which is considered to be a plane wave in our model. Both of these simplifications may lead to slightly overestimated values of the optical $Q$ factor.

\subsection{\label{Section_S2_subst} Impact of the substrate}

We first study the DPhoC structure without substrate. This structure has a mirror symmetry plane in the middle between the two photonic-crystal membranes.
As previously established in Ref.~\cite{Fitzgerald2021}, it exhibits a bound state in the continuum, as can be seen from the characteristic shape of the quality factor as a function of hole radius shown in Fig.~\ref{Q(subst,abs,r)}(a) without material absorption, $\mathrm{Im}(n)=0$ (blue markers). While the quality factor is about $1\times10^4$ at $r$ = 400\,nm, it exponentially increases up to $2\times10^{9}$ at $r$ = 418.1\,nm. This value keeps increasing without limit close to the BIC condition. Introducing dissipative loss from material absorption [orange markers in Fig.~\ref{Q(subst,abs,r)}(a)], the BIC resonance is turned into a dissipative-loss-limited quasi-BIC with a maximum quality factor of about $7\times10^5$.

\begin{figure}
    \includegraphics{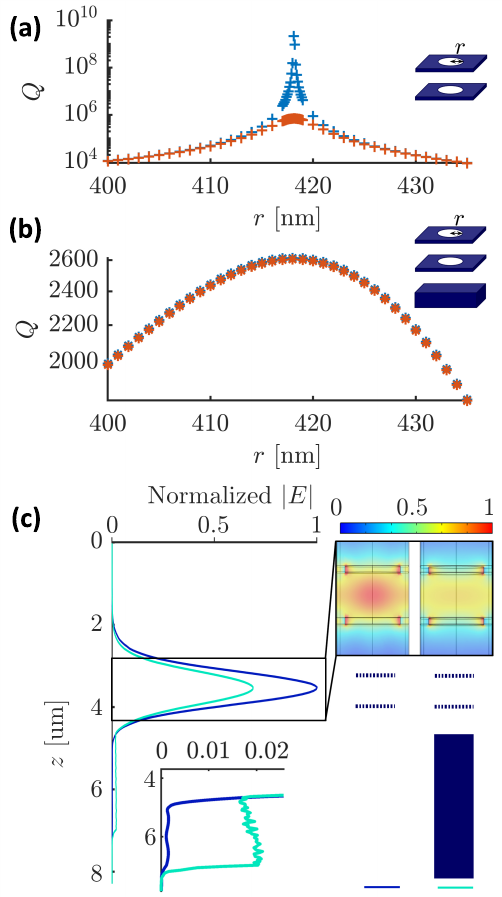}
    \caption{\label{Q(subst,abs,r)} Cavity quality factor as a function of hole radius without (blue markers) and with (orange markers) absorption in GaAs, for the DPhoC (a) and DPhoCS (b). (c) Electric field strength along the $z$ axis, with $x$ and $y$ taken at the center of the unit cell (left) and field distribution in the $yz$ plane (right), for the DPhoC (dark blue line) and DPhoCS (cyan line). The field is normalized by the energy in the cavity. $a$ = 1085\,nm, $r$ = 418\,nm, and $q$ = 680\,nm.}
\end{figure}

In an experiment, such a double membrane would be fabricated on top of a substrate. We therefore introduce a GaAs substrate (DPhoCS system) and calculate the quality factor $Q$ under the exact same conditions as for the DPhoC [see Fig.~\ref{Q(subst,abs,r)}(b)].  
Comparing the quality factors presented in Figs.~\ref{Q(subst,abs,r)}(a) and \ref{Q(subst,abs,r)}(b), we observe that adding the substrate causes $Q$ to drop from over $10^5$ to about $2.6\times10^3$. Even without absorption, the quality factor remains finite in the presence of a substrate, demonstrating that adding the substrate destroys the bound state in the continuum. Light can leak out from the cavity through the substrate, as shown by the propagating field from $z$ = 4.5\,$\mu$m [cyan line in Fig.~\ref{Q(subst,abs,r)}(c)], while in the case of the DPhoC the field is evanescent around the cavity (dark blue line).

\begin{figure}
    \includegraphics{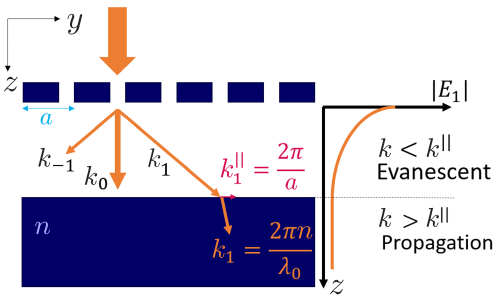}
    \caption{\label{Diffraction_sq} Left: Schematic representation of the zeroth and first orders of diffraction from a 2D PhC slab patterned with a square lattice. Right: Representation of the electric field strength for the first order of diffraction along the $z$ axis.}
\end{figure}

This effect can be explained by the generation of higher orders of diffraction in the substrate, illustrated in Fig.~\ref{Diffraction_sq}. Without the substrate present, the structure is bounded by air and only the zeroth diffraction order can propagate out of the structure for subwavelength photonic-crystal slabs with $a<\lambda_0$; all higher orders are evanescent since their wavevector's parallel component is larger than their wavevector's magnitude ($k_m^{||}>k_m$ with $m\neq0$). However, this is no longer the case in materials with a high refractive index such as GaAs. In the presence of the substrate, the first diffraction order is evanescent in airwith $k_1=2\pi/\lambda_0<k_{1}^{||}$, but propagates in the high-$n$ substrate with $k_1=2\pi n/\lambda_0>k_{1}^{||}$ since $a>\lambda_0/n$. The closer the substrate is to the PhC, the more energy is leaked away through the diffraction channel. This is shown in Fig.~\ref{Q(dcs,r)}(a), where the distance between the cavity and the substrate, $d_\mathrm{cs}$, is increased up to three times its original value $q$. As the substrate gets further away from the cavity, less energy can leak from the cavity by evanescent coupling to the higher diffraction orders. Considering constant dissipative losses and reducing radiative losses with increasing $d_\mathrm{cs}$, with large enough distance $d_\mathrm{cs}$, the radiative losses through the substrate become smaller than the dissipative losses and a quasi-BIC is recovered, allowing one to restore $Q$ to its value observed with no substrate present (DPhoC). The field distribution confirms this assumption, as the energy leaking out of the cavity decreases as the distance to the substrate increases [from dark blue to cyan lines in Fig.~\ref{Q(dcs,r)}(b)]. This result highlights how much $d_\mathrm{cs}$, and thus the whole system, would need to be enlarged in order to recover a quasi-BIC. Such larger systems are not desirable and, thus, we consider an alternative solution in the following.

\begin{figure}
    \includegraphics{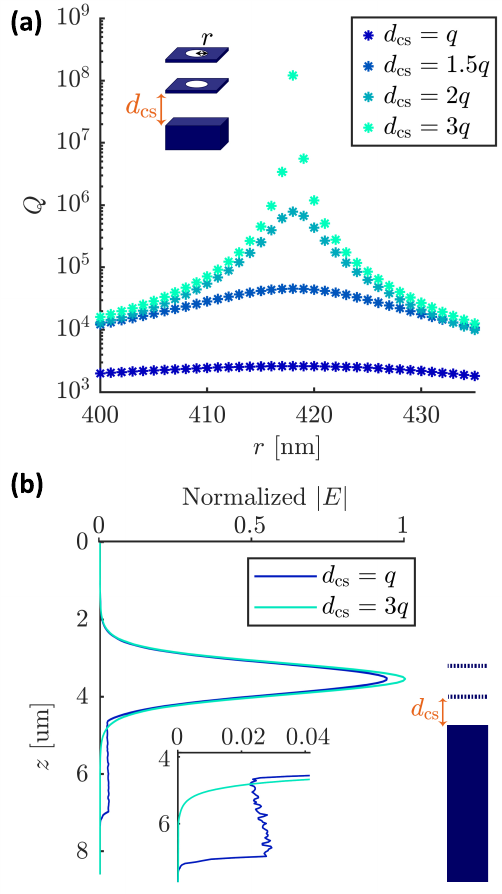}
    \caption{\label{Q(dcs,r)} (a) Quality factor of the DPhoCS as a function of hole radius, for different distances $d_\mathrm{cs}$ between the bottom PhC and the substrate. No material absorption is modeled. (b) Electric field strength along the $z$ axis , with $x$ and $y$ taken at the center of the unit cell, for the DPhoCS, for different distances $d_\mathrm{cs}$ between the bottom slab and the substrate. The field is normalized by the energy in the cavity. $a$ = 1085\,nm, $r$ = 418\,nm, and $q$ = 680\,nm. No material absorption is modeled.}
\end{figure}

\subsection{Single photonic-crystal membrane above a perfect mirror}

The fabrication of the DPhoC system with its identical PhC slabs requires high precision during the etching process and growth, especially since the BIC is very sensitive to small variations between the patterning of both PhC slabs~\cite{Fitzgerald2021}. 
We can, however, make use of the mirror plane symmetry in the DPhoC system. Instead of having a second PhC, we imagine a single membrane at a distance $q/2$ over a perfect mirror, modeled with a perfectly magnetic conductor (PMC) boundary condition. Effectively, this surface mirrors the PhC membrane and distances in air. Thus, a distance $q/2$ to a perfect mirror corresponds to the $q$-long cavity considered before. As demonstrated by Refs.~\cite{hsu2013bloch,yu2021ultra}, a Fabry-Perot-type BIC can be realized in such systems.

Using the same parameters ($q$ = 680\,nm, $a$ = 1085\,nm) above a perfect mirror at a distance $q/2$, the quality factor as a function of the hole radius is calculated. Fig.~\ref{Q(S2_SLPMC,r)} compares the resulting quality factor with the one previously obtained for the DPhoC system with no material absorption. We observe that both systems demonstrate the same values of $Q$.
As shown in Fig.~\ref{SL DBR_PEC_Q(syst,q)} (cross and triangle markers), varying $q$ rather than $r$ also results in the formation of a BIC at the exact same parameters. As an implementation of the perfect mirror, one could use a distributed Bragg reflector (DBR) stack. We choose a DBR because of its low absorption properties in the telecom range, simple design, and compatibility with the growth and fabrication process of optomechanical microcavities. Such optomechanical cavities consisting of a mechanically compliant periodic grating facing a DBR mirror have been proposed in Refs.~\cite{naesbyMicrocavitiesSuspendedSubwavelength2018,cernotikCavityQuantumElectrodynamics2019} in the context of optomechanical linewidth narrowing, and experimentally realized in Refs.~\cite{xu2022millimeter,zhouCavityOptomechanicalBistability2022}. The turquoise dots in Fig.~\ref{SL DBR_PEC_Q(syst,q)} represent the quality factor obtained with this system consisting of one PhC membrane and a DBR (denoted SL-DBR), which we will further study in the next section. Unlike an infinitely thin perfect mirror, all the light is not immediately reflected at the surface of a DBR. This leads to a BIC appearing at a larger $q$, as seen by the small misalignment of the maximum $Q$ values between the SL-DBR system and the two other systems.

\begin{figure}
    \includegraphics{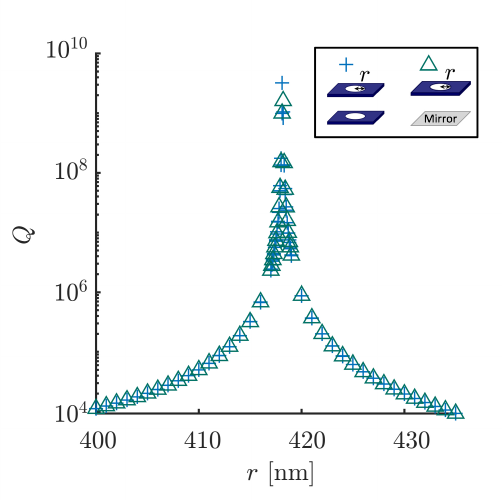}
    \caption{\label{Q(S2_SLPMC,r)} Quality factor as a function of hole radius, for the DPhoC system (crosses) and a system consisting of one PhC membrane facing a perfect mirror (triangles). No material absorption is modeled. $q$ = 680\,nm and $a$ = 1085\,nm.}
\end{figure}

\begin{figure}
    \includegraphics{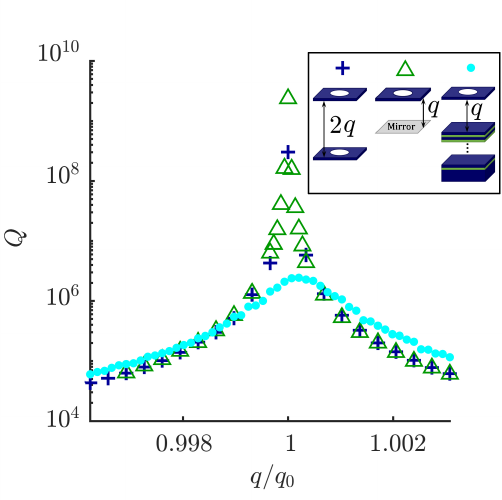}
    \caption{\label{SL DBR_PEC_Q(syst,q)} Quality factor of DPhoC with a cavity length $2q$ (crosses), of a single PhC membrane and a perfect mirror with a cavity length $q$ (triangles), and of SL-DBR with a cavity length $q$ (dots). No material absorption is modeled. Insets are schematic representations of each system. $q_0$ is used for normalization and is taken at the BIC condition for DPhoC.}
\end{figure}

\subsection{Shifting the BIC frequency}

\begin{figure}
    \includegraphics{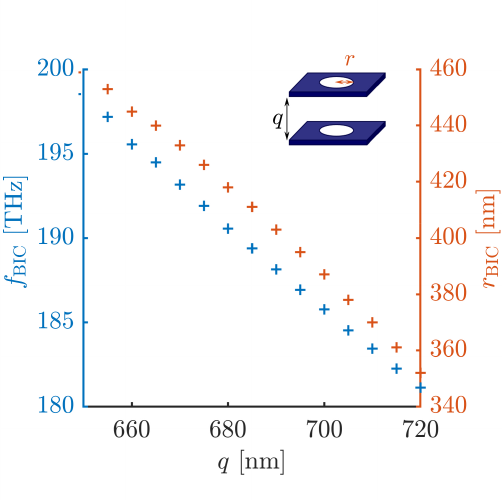}
    \caption{\label{r&f_BIC(q)} BIC frequency $f_\mathrm{BIC}$ (blue markers) and hole radius $r_\mathrm{BIC}$ (orange markers) for the DPhoC with varying cavity lengths $q$ for $a$ = 1085\,nm.}
\end{figure}

The above results demonstrate that a quasi-BIC can be found in a microcavity on a substrate for a specific set of parameters ($q$ = 680\,nm, $a$ = 1085\,nm) at $r_\mathrm{BIC}$ = 418.1\,nm and $f_\mathrm{BIC}$ = 190.5\,THz. In experiments, it is useful to have control over the resonance frequency of the BIC. This can be achieved by controlling the structural parameters of the microcavity, e.g., the cavity length $q$. 
In Fig.~\ref{r&f_BIC(q)}, we show how the BIC frequency can be tuned inside our frequency range of interest (see Sec.~\ref{sec: Model description}) with variations of $q$ (blue markers). Longer cavities have smaller $f_\mathrm{BIC}$: this is expected as larger structures have lower resonance frequencies. However, these variations on the frequency cannot be achieved without adjusting the PhC parameters, as these parameters affect the PhC resonance. For example, we consider here variations of $r$, while keeping $a$ and $e$ constant (orange markers). For the considered mode, these decreased $f_\mathrm{BIC}$ are realized by reduced $r_\mathrm{BIC}$.

\section{\label{Section_SL-DBR}Photonic-crystal membrane above a Bragg mirror}

We now proceed by implementing the perfect mirror in the model using a DBR. Here we have to remember from the previous analysis that the photonic-crystal membrane creates higher orders of diffraction inside the cavity. To prevent the first order from leaking out into the high-index substrate---creating a radiative loss channel and destroying the BIC---we need a highly reflective mirror not only for the zeroth but also for the first order of diffraction. Therefore, the DBR should be specifically designed to reflect these two orders of diffraction, represented in Fig.~\ref{Diffraction_sq}. Here we focus on only the first two orders of diffraction. Even higher orders of diffraction may also propagate, but their influence is negligible as they carry less energy.

This brings us to the system depicted in Fig.~\ref{syst}(b), a cavity consisting of a single photonic-crystal membrane above a DBR grown on a GaAs substrate.

We design the cavity to have a resonance wavelength at $\lambda_0$ = 1525\,nm. A highly reflective Fano resonance of the PhC is obtained at $\lambda_0$ for $a$ = 1081\,nm, $r$ = 457.5\,nm, and $e$ = 109\,nm. The membrane can be seen as a periodic grating that diffracts the incoming light mostly in the zeroth and first orders of diffraction. These two modes should then be reflected by the DBR. The DBR is made of $N$ periods of alternating layers of GaAs and Al$_x$Ga$_{1-x}$As of thicknesses $e_\mathrm{GaAs}$ and $e_\mathrm{AlGaAs}$, respectively. For the current model we use aluminum content $x$ = 0.97.
A DBR with two layers in the unit cell allows to achieve a bandgap for two different parallel momenta. Based on the equations of the DBR angle-dependant reflectivity from Ref.~\cite{Yariv1984}, a homemade optimization algorithm was used to calculate the values $e_\mathrm{GaAs}$ and $e_\mathrm{AlGaAs}$. With the refractive index of AlGaAs set to $n_\mathrm{AlGaAs}$ = 2.927, a high reflectivity of the DBR for both the zeroth and first orders of diffraction is achieved when $e_\mathrm{GaAs}$ = 100.2\,nm and $e_\mathrm{AlGaAs}$ = 147.1\,nm. We find that 40 periods are sufficient to suppress the radiative loss channel through the substrate to recover the quasi-BIC.

\subsection{Near-field effects}

\begin{figure}
    \includegraphics{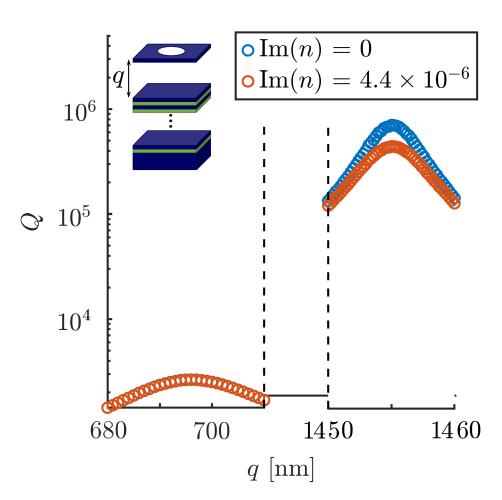}
    \caption{\label{SLDBR_Q(abs,q)} Cavity quality factor as a function of cavity length without (blue markers) and with material absorption (orange markers), for the SL-DBR. The cavity is formed by a GaAs PhC slab (air hole radius $r$ = 457.5\,nm, $a = 1081$\,nm) arranged in a square lattice, and a DBR consisting of 40 periods of alternating GaAs and AlGaAs layers with thicknesses of 100.2\,nm and 147.1\,nm, respectively.}
\end{figure}

\begin{figure}
    \includegraphics{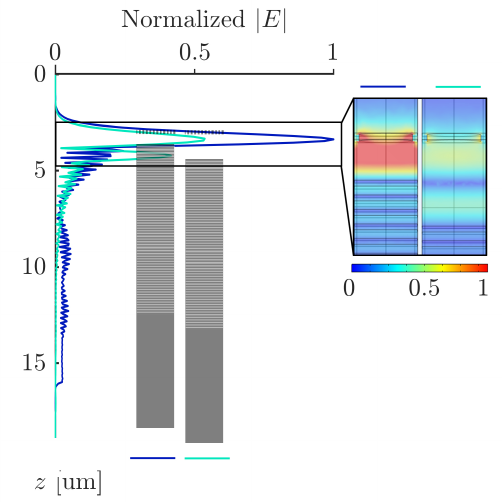}
    \caption{\label{E_SLDBR_E(q)} Electric field strength along the $z$ axis, with $x$ and $y$ taken at the center of the unit cell, for the SL-DBR, for $q$ = 696\,nm (dark blue line) and $q_\mathrm{l}$ = 1455\,nm (cyan line). Insets are electric field distributions in a $yz$ plane. The field is normalized by the energy in the cavity. The cavity is formed by a 2D PhC slab of cylindrical holes (radius $r$ = 457.5\,nm) arranged in a square lattice with a lattice constant of 1081\,nm surrounded by GaAs, and a DBR consisting of 40 periods of alternating GaAs and AlGaAs layers with a thickness of 100.2\,nm and 147.1\,nm, respectively.}
\end{figure}

It is crucial to understand the influence of the DBR proximity to the PhC membrane and the near-field effects associated with it when choosing the BIC-based microcavity parameters.
The quality factor of the microcavity is shown in Fig.~\ref{SLDBR_Q(abs,q)} for two ranges of the cavity length: from 680 to 710\,nm and from 1450 to 1460\,nm.
For the shorter cavity (left part of the plot), the maximum of the quality factor is limited to $2.6\times10^3$. In cases without and with absorption the $Q$ values are almost identical. Our calculations indicate that the impact on $Q$ from material absorption is minor compared to the evanescent coupling, as the considered DBR is not designed to reflect the near-field evanescent wave. In these shorter cavities, for which the PhC and the DBR are in close proximity, this near-field evanescent wave from the PhC membrane can reach the DBR. Light then propagates through the DBR and reaches the substrate [dark blue line in Fig.~\ref{E_SLDBR_E(q)}], as shown by the oscillations in the DBR (from $z$ = 4\,$\mu$m to $z$ = 14\,$\mu$m) and then its propagation in the substrate.
A longer cavity alleviates that issue: with increased distance, this near-field evanescent wave cannot reach the DBR anymore. Increased cavity lengths may thus be a solution to avoid leaking through evanescent coupling. When the cavity length is increased by an integral number of half wavelengths, the BIC reappears~\cite{Fitzgerald2021}. We can thus observe another peak of the quality factor at $q_\mathrm{l}$ = 1455\,nm. In this case, the near-field coupling to the substrate is significantly reduced and $Q_\mathrm{max}$ reaches $7.1\times10^5$ and $4.4\times10^5$ without and with material absorption, respectively. For a $q_\mathrm{l}$-cavity, the electric field strength in the DBR decreases exponentially (light blue line in Fig.~\ref{E_SLDBR_E(q)}). For an even longer cavity ($q$ = $q_l+\lambda_0/2$ = 2214\,nm), $Q$ reaches $1.2\times10^6$ when no absorption is modeled. This second improvement on $Q$ seems relatively low, proving that most evanescent coupling has been suppressed by increasing the cavity length by just one-half wavelength. 

It is worth noting that even at larger $q$ and without material absorption, the quality factor remains finite, a sign that some radiative loss channel still exists in the system, namely radiation into the substrate because of the finite reflectivity of the DBR. This is not problematic, however, since we can make the radiative losses smaller than the dissipative losses.

In order to avoid these near-field effects, we will further only consider longer cavities ($q \approx$ 1455\,nm).

\subsection{Photonic-crystal membrane with hexagonal lattice}
\label{Section_hexa}

As the 2D pattern of the PhC can be freely modified, we can examine different symmetries. In this section, we study a hexagonal lattice. The reciprocal lattice is then also hexagonal and the magnitude of its primitive lattice vectors is $G=4\pi/(a\sqrt{3})$.
The first diffraction order from a membrane with a hexagonal lattice propagates with an in-plane wave vector $k_{||}=G$~\cite{ochiai2001dispersion,barnes2004surface}. For the same lattice constant $a$, this value is about 15\% larger than that for a square lattice for which $k_{||} = 2\pi/a$ and the diffraction angle from a hexagonal lattice is therefore larger than that from a square lattice. It also means that the cut-off for the first diffraction order is higher for a hexagonal lattice. We thus expect the impact from the higher orders of diffraction on the quality factor to be lower for a hexagonal lattice.

This is indeed what we observe if the PhC pattern in the SL-DBR system is changed from a square to a hexagonal lattice. This modification is accompanied by an adjustment of the 2D PhC parameters in order to track the BIC resonance: while keeping the DBR parameters, and $a$ and $e$ constant, $r_\mathrm{hex}$ is adjusted to 225\,nm.

\begin{figure}
    \includegraphics{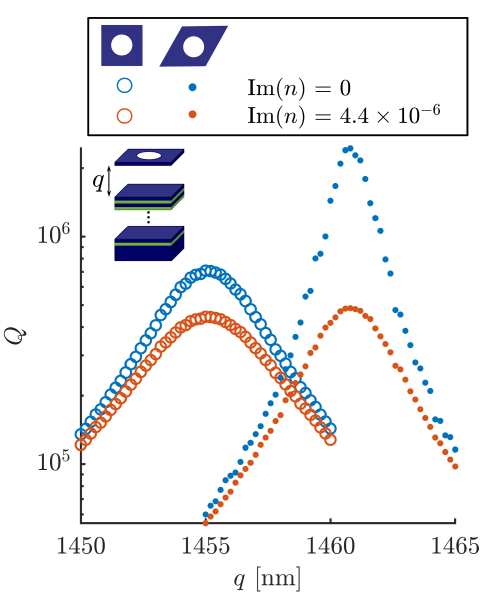}
    \caption{\label{Q(latt,abs)} Cavity quality factor as a function of cavity length $q$ without (blue markers) and with material absorption (orange markers), for a square (sq) (empty round markers) and hexagonal (hex) lattices (dot markers), for the SL-DBR. The same DBR ($N$ = 40, $e_\mathrm{GaAs}$ = 100.2\,nm, $e_\mathrm{AlGaAs}$ = 147.1\,nm) is used in all cases. The 2D PhC has $a$ = 1081\,nm and $e$ = 109\,nm. $r_\mathrm{sq}$ is 457.5\,nm and $r_\mathrm{hex}$ is 225\,nm.}
\end{figure}

A quality factor comparison between square and hexagonal lattices is shown in Fig.~\ref{Q(latt,abs)}. Only the hole radius is adjusted to obtain the resonance of the 2D PhC slab. The hole radii for the square and hexagonal lattices are denoted by $r_\mathrm{sq}$ and $r_\mathrm{hex}$, respectively.\\
Looking first at the case without material absorption (blue markers), $Q$ reaches a maximum value of about $7.1\times10^5$ for a square lattice, whereas a maximum $Q$ of $2.4\times10^6$ is reached for the hexagonal lattice, in agreement with our earlier prediction that the quality factor should increase for a hexagonal lattice due to the lesser impact of the higher orders of diffraction. On the other hand, when comparing the case with material absorption (orange markers), the difference between square and hexagonal lattices becomes much smaller. Their maximum values are very close: about $4.4\times10^5$ for the square lattice and $4.8\times10^5$ for the hexagonal lattice. Since the resonances are dissipative-loss-limited, further reducing the radiative losses with a hexagonal lattice has limited impact.

We note that the DBR layer thicknesses were originally optimized to make the DBR reflect both the zeroth and first orders of diffraction coming from the 2D PhC patterned with a square lattice. While the zeroth order of diffraction remains normally incident on the DBR, the angle for the first order of diffraction depends on $k_{||}$. However, changing from a square lattice to a hexagonal lattice modified $k_{||}$ and the DBR may no longer reflect this first order of diffraction optimally.

\begin{figure}
    \includegraphics{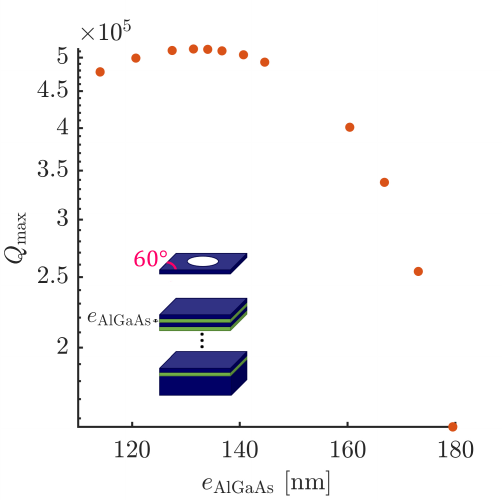}
    \caption{\label{Q(eDBR)} Cavity quality factor as a function of the DBR AlGaAs layers when the DBR GaAs layer thickness is adjusted to maximize $Q$. The cavity length is set to 1460.8\,nm. Material absorption is included in the calculations.}
\end{figure}

Therefore, a new optimization of the DBR layers thicknesses is necessary. For each value of $e_\mathrm{GaAs}$, we evaluated $Q$ as a function of $e_\mathrm{AlGaAs}$ and extracted the maximum, $Q_\mathrm{max}$. Then we repeated this process for other values of $e_\mathrm{GaAs}$ and plotted $Q_\mathrm{max}$ as a function of $e_\mathrm{AlGaAs}$, shown in Fig.~\ref{Q(eDBR)}. The highest $Q_\mathrm{max}$ obtained in this way is about $5.1\times10^5$, for $e_\mathrm{GaAs}$ = 112\,nm and $e_\mathrm{AlGaAs}$ = 131.4\,nm. This study includes a realistic material absorption for GaAs.

\section{Conclusion}
We have proposed an optomechanical microcavity supporting a photonic quasi-BIC, which consists of a PhC membrane suspended above a DBR mirror. We found that for the realization of a quasi-BIC it is essential to design the DBR mirror such that it reflects both the normally incident wave and the first diffraction order. When implementing such a cavity with a GaAs PhC membrane \cite{kinimanjeshwarSuspendedPhotonicCrystal2020}, it is predicted to yield optical quality factors of $5\times10^5$. This value is limited by both material absorption and radiative loss channels through the DBR. Our results could be found by generalizing other numerical or semianalytical methods, e.g., by modeling the reflection of our cavity as an effective surface impedance~\cite{hessel1965,tassin2012b,monticone2017} or using an entire domain integral equation analysis~\cite{tsitsas2007,tsitsas2017}. At a later stage, to improve the results in comparison to experimental measurements, our analysis can be further extended by considering the finite size and defects in the periodicity of the PhC membrane, which are both not currently captured by our periodic boundary condition. These effects lead to small variations in the resonance frequency~\cite{zundel2018finite} and reduction of the total reflectivity \cite{toft-vandborgCollimationFinitesizeEffects2021}.  Furthermore, we currently model the incident beam as a plane wave. A realistic Gaussian beam could be modeled by a superposition of plane waves of different amplitudes and incident angles~\cite{yang2005two,kinimanjeshwarSuspendedPhotonicCrystal2020,toft-vandborgCollimationFinitesizeEffects2021}. Finally, GaAs is an optically isotropic material, resulting in polarization-independent reflection off the SL-DBR cavity. When other materials are used for the membrane, an in-plane anisotropy may lead to interesting polarization behavior~\cite{valagiannopoulos2017manipulating}, such as a polarization-dependent BIC frequency. By carefully controlling the polarization of the incident light, this system would then be able to allow for two orthogonal optical and resonant modes, offering more control possibilities.

The predicted optical quality factors would result in a cavity decay rate of $\kappa\approx2\pi\times 100\,$MHz for a microcavity of $L\approx1.55\,\mu$m length. Typical mechanical frequencies of suspended PhC membranes lie in the kHz to MHz range with an effective mass of a few nanograms \cite{bui2012high,makles20152d,norte2016mechanical,kinimanjeshwarSuspendedPhotonicCrystal2020,zhouCavityOptomechanicalBistability2022,manjeshwarMicromechanicalHighQTrampoline2022}. When using such PhC membranes as an end-mirror \cite{zhouCavityOptomechanicalBistability2022} in our proposed microcavity, an optomechanical frequency pulling factor of about $2\pi\times100\,$GHz/nm would be realized, leading to a single-photon coupling strength on the order of $2\pi\times500\,$kHz and a ratio of coupling strength to optical loss of $g_0/\kappa\sim5\times 10^{-3}$, which is comparable to realizations based on optomechanical crystals \cite{chanLaserCoolingNanomechanical2011}, microwave optomechanics \cite{teufelCircuitCavityElectromechanics2011}, or magnetomechanics \cite{rodriguesCouplingMicrowavePhotons2019,schmidtSidebandresolvedResonatorElectromechanics2020,zoepflKerrEnhancedBackaction2023}. Our proposed quasi-BIC optomechanical microcavity would be placed in the non-sideband resolved regime, which is amenable for optomechanical feedback cooling \cite{genesGroundstateCoolingMicromechanical2008,rossiMeasurementbasedQuantumControl2018} and sensing \cite{liCavityOptomechanicalSensing2021}. A further increase of the optical quality factor of the quasi-BIC microcavity could be realized by designing DBRs that reflect even higher orders of diffraction, considering materials with lower absorption, or employing machine-learning based optimization to find PhC patterns with reduced dissipative loss~\cite{jiangDeepNeuralNetworks2021,kudyshevMachineLearningIntegrated2021,gahlmannDeepNeuralNetworks2022}. 

\begin{acknowledgements}
This work was supported in part by the \mbox{QuantERA} project C’MON-QSENS!, by the Knut and Alice Wallenberg Foundation through a Wallenberg Academy Fellowship (W.W.), by the Wallenberg Center for Quantum Technology (WACQT, A.C.), by Chalmers Area of Advance Nano, and by the Swedish Research Council (Grant No.~2019-04946 and No.~2020-05284). Calculations were  performed on resources provided by the Swedish National Infrastructure for Computing (NAISS), at the KTH/PDC site, partially funded by the Swedish Research Council under Grant No. 2022-06725.
\end{acknowledgements}

\appendix

\section{Model parameters}

\begin{figure}
    \includegraphics{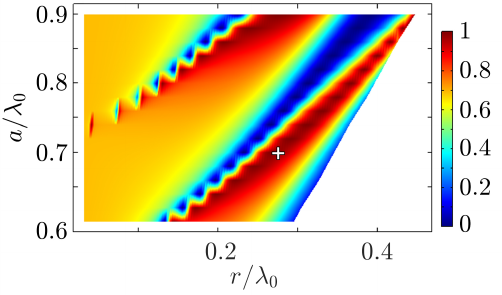}
    \caption{\label{SL_R(a,r)} Reflectance of a single 99\,nm-thick photonic-crystal membrane made of GaAs patterned with cylindrical holes, as a function of $r$ and $a$, for a frequency $f_0$ = 190 THz. No material absorption is modeled.}
\end{figure}

\begin{table}
\caption{\label{Param}Default parameters used in the simulation, for each modeled system.}
\begin{tabular}{c||c|c|c}
 & DPhoC(S) & SL-DBR (sq) & SL-DBR (hex) \\ [0.5ex]
\hline \hline
$a$ [nm]& 1085& 1081 & 1081\\
\hline
$r$ [nm] & 418 & 457.5 & 225\\
\hline
$q$ [nm] & 680 & 1455 & 1461\\
\hline
$e$ [nm] & 99 & 109 & 109\\
\hline
$e_\mathrm{GaAs}$ [nm] & / & 100.2 & 112\\
\hline
$e_\mathrm{AlGaAs}$ [nm] & / & 147.1 & 131.4\\
\hline
$N$ & / & 40 & 40\\
\hline
$d_\mathrm{cs}$ [nm] & 680 & / & /\\
\end{tabular}
\end{table}

Unless otherwise stated, Table~\ref{Param} summarizes all parameters used in the simulations. These are optimized values, leading to the highest $Q$ obtained in our simulations for each considered system. Starting parameters such as $a$ and $r$ were set at a resonance of a single PhC membrane. To define this, the reflectivity of a single membrane is calculated and shown in a 2D map (Fig.~\ref{SL_R(a,r)}). A white cross is placed at the chosen parameters. A larger resonance was chosen to account for possible inaccuracies during the fabrication process. Other parameters, such as the cavity length, are then obtained through numerical optimization. The resulting optimized values are all gathered in Table~\ref{Param} for the different cavities discussed in this article. 

\section{Band diagrams}

\begin{figure}[b]
    \includegraphics{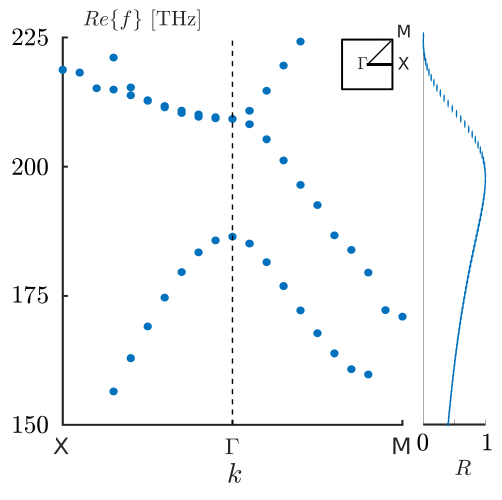}
    \caption{\label{sq_bands} Band diagram (left) and reflectance spectra at the $\Gamma$ point (right) of a 109-nm-thick photonic-crystal membrane made of GaAs patterned with cylindrical holes ($r$ = 457.5\,nm) in a square lattice ($a$ = 1081\,nm).}
\end{figure}

\begin{figure}
    \includegraphics{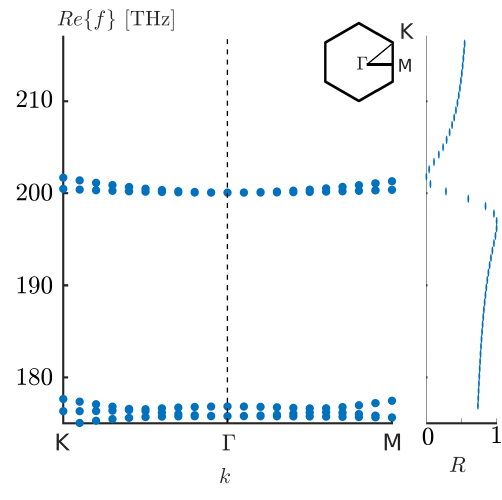}
    \caption{\label{hex_bands} Band diagram (left) and reflectance spectra at the $\Gamma$ point (right) of a 109-nm-thick photonic-crystal membrane made of GaAs patterned with cylindrical holes ($r$ = 225\,nm) in a hexagonal lattice ($a$ = 1081\,nm).}
\end{figure}

It is worth noting that a hexagonal lattice is a lot less sensitive to variations of the incidence angle than a square lattice. This is shown by the band diagrams of the single photonic-crystal slab patterned with a square and hexagonal lattice in Figs.~\ref{sq_bands} and \ref{hex_bands}, respectively. Bands for the hexagonal lattice appear flat compared to those for the square lattice.\\
A reflectance spectrum at the $\Gamma$ point is aligned at the right, proving agreement between these results on the $\Gamma$ point, as a high reflectance is observed in photonic bandgaps.

%

\end{document}